\documentstyle[prd,aps,preprint,tighten,epsfig]{revtex}

\begin{document}

\draft

\title{Right Unitarity Triangles, Stable CP-violating Phases
and Approximate Quark-Lepton Complementarity}
\author{{\bf Zhi-zhong Xing}
\thanks{E-mail: xingzz@ihep.ac.cn}}
\address{Institute of High Energy Physics
and Theoretical Physics Center for Science Facilities, \\ Chinese
Academy of Sciences, P.O. Box 918, Beijing 100049, China}
\maketitle

\begin{abstract}
Current experimental data indicate that two unitarity triangles of
the CKM quark mixing matrix $V$ are almost the right triangles with
$\alpha \approx 90^\circ$. We highlight a very suggestive
parametrization of $V$ and show that its CP-violating phase $\phi$
is nearly equal to $\alpha$ (i.e., $\phi - \alpha \approx
1.1^\circ$). Both $\phi$ and $\alpha$ are stable against the
renormalizaton-group evolution from the electroweak scale $M^{}_Z$
to a superhigh energy scale $M^{}_X$ or vice versa, and thus it is
impossible to obtain $\alpha = 90^\circ$ at $M^{}_Z$ from $\phi =
90^\circ$ at $M^{}_X$. We conjecture that there might also exist a
maximal CP-violating phase $\varphi \approx 90^\circ$ in the MNS
lepton mixing matrix $U$. The approximate quark-lepton
complementarity relations, which hold in the standard
parametrizations of $V$ and $U$, can also hold in our particular
parametrizations of $V$ and $U$ simply due to the smallness of
$|V^{}_{ub}|$ and $|V^{}_{e3}|$.
\end{abstract}

\pacs{PACS number(s): 12.15.Ff, 12.38.Bx, 12.10.Kt, 13.10.+q,
14.60.Pq, 25.30.Pt}

\newpage

\section{Introduction}

In the standard model (SM) of electroweak interactions, it is the
$3\times 3$ Cabibbo-Kobayashi-Maskawa (CKM) matrix that provides
an elegant and consistent description of the observed phenomena of
quark flavor mixing and CP violation \cite{CKM}. Unitarity is the
only but powerful constraint, imposed by the SM itself, on the CKM
matrix $V$. This constraint can be expressed as two sets of
orthogonality-plus-normalization conditions:
\begin{equation}
\sum_\alpha \left( V^{}_{\alpha i} V^*_{\alpha j} \right) \; =\;
\delta^{}_{ij} \; , ~~~~ \sum_i \left( V^{}_{\alpha i} V^*_{\beta
i } \right) \; =\; \delta^{}_{\alpha\beta} \; ,
\end{equation}
where the Greek subscripts run over the up-type quarks $(u,c,t)$
and the Latin subscripts run over the down-type quarks $(d,s,b)$.
The six orthogonality relations correspond to six triangles in the
complex plane, the so-called unitarity triangles. Among them
\footnote{Here we follow Ref. \cite{FX00} to name each CKM unitarity
triangle by using the flavor index that does not manifest in its
three sides.},
the triangle $\triangle^{}_s$ is most popular because both its
three inner angles (defined as $\alpha$, $\beta$ and $\gamma$ in
Fig. 1) and its three sides can well be determined at the
$B$-meson factories \cite{PDG08}. The counterpart of
$\triangle^{}_s$ is the unitarity triangle $\triangle^{}_c$ (as
shown in Fig. 1), which will be measured and reconstructed at the
LHC-$b$ \cite{LHCb} and (or) the super-$B$ factory \cite{SB}. Note
that one of the inner angles of $\triangle^{}_c$ is equal to the
inner angle $\alpha$ of $\triangle^{}_s$. Current experimental
data \cite{PDG08} tell us that these two triangles are
approximately congruent with each other. For example, the inner
angles $\xi$ and $\zeta$ of $\triangle^{}_c$ are very close to
$\beta$ and $\gamma$ of $\triangle^{}_s$:
\begin{equation}
\xi - \beta \; \approx \; \gamma - \zeta \; \approx \; \lambda^2
\eta \; \approx \; 1^\circ \; ,
\end{equation}
where $\lambda \approx 0.226$ and $\eta \approx 0.35$ are the
well-known Wolfenstein parameters \cite{W} in an ${\cal
O}(\lambda^4)$-expansion of the CKM matrix $V$ \cite{Xing95}. A
more striking result is $\alpha \approx 90^\circ$ obtained by the
CKMfitter Group \cite{CKMfit} and the UTfit Collaboration
\cite{UTfit}. If $\alpha = 90^\circ$ holds exactly, then both
$\triangle^{}_s$ and $\triangle^{}_c$ will be the right triangles.

\vspace{0.4cm}

The possibility of $\alpha \approx 90^\circ$ was actually
conjectured a long time ago in an attempt to explore the realistic
texture of quark mass matrices \cite{FX95}, and it has recently
been remarked from some different phenomenological points of view
\cite{Koide,Masina,Harrison}. Here we are interested in the
following questions and possible answers to them:
\begin{itemize}
\item     What is the immediate consequence of $\alpha = 90^\circ$ on the
CKM matrix $V$ and its four independent parameters?

\item     Could $\alpha \approx 90^\circ$ result from an underlying but more
fundamental CP-violating phase $\phi = 90^\circ$ in the quark mass
matrices or in the CKM matrix?

\item     Is the result $\alpha \approx 90^\circ$ or $\phi =
90^\circ$ stable against quantum corrections, for instance, from
the electroweak scale $M^{}_Z$ to a superhigh-energy scale
$M^{}_X$ (such as the scale of grand unified theories or the scale
of neutrino seesaw mechanisms)? In other words, is $\alpha =
90^\circ$ at $M^{}_Z$ possibly a natural low-energy consequence of
$\phi = 90^\circ$ at $M^{}_X$ due to the renormalization-group
running effect?

\item     Could the $3\times 3$ Maki-Nakagawa-Sakata (MNS) neutrino mixing
matrix $U$ \cite{MNS} contain a similar maximal CP-violating phase
$\varphi = 90^\circ$?
\end{itemize}
We shall point out that $\alpha = 90^\circ$ simply implies ${\rm
Re}(V^{}_{tb}V^{}_{ud}V^*_{td}V^*_{ub}) = 0$. Given a very
suggestive parametrization of $V$ advocated in Ref. \cite{FX97},
we show that its CP-violating phase $\phi$ is nearly equal to
$\alpha$; i.e., $\phi - \alpha \approx 1.1^\circ$. But we find
that both $\phi$ and $\alpha$ are rather stable in the
renormalizaton-group evolution from $M^{}_Z$ up to $M^{}_X$ or
vice versa, and thus it is impossible to obtain $\alpha =
90^\circ$ at $M^{}_Z$ from $\phi = 90^\circ$ at $M^{}_X$ by
attributing the tiny difference $\phi - \alpha \approx 1.1^\circ$
to radiative corrections. We shall briefly discuss the approximate
quark-lepton complementarity relations both in the standard
parametrizations of $V$ and $U$ and in our particular
parametrizations of $V$ and $U$, and then make a conjecture of the
maximal CP-violating phase $\varphi \approx 90^\circ$ for the MNS
matrix $U$ at the end of this paper. We hope that some of our
points, which might be helpful for building phenomenological
models, can soon be tested with more accurate experimental data on
quark and lepton flavor mixing parameters.

\section{Implications of $\alpha = 90^\circ$}

Let us define the Jarlskog invariant of CP violation ${\cal
J}^{}_q$ for the CKM matrix $V$ \cite{J}:
\begin{equation}
{\rm Im}\left( V^{}_{\alpha i} V^{}_{\beta j} V^*_{\alpha j}
V^*_{\beta i} \right) \; =\; {\cal J}^{}_q \sum_{\gamma}
\epsilon^{}_{\alpha\beta\gamma} \sum_{k} \epsilon^{}_{ijk} \; ,
\end{equation}
where the Greek and Latin subscripts run over $(u,c,t)$ and
$(d,s,b)$, respectively. All six unitarity triangles of $V$ have
the same area which amounts to ${\cal J}^{}_q/2$. Triangles
$\triangle^{}_s$ and $\triangle^{}_c$ in Fig. 1 correspond to the
orthogonality relations
\begin{eqnarray}
V^{}_{ud} V^*_{ub} + V^{}_{cd} V^*_{cb} + V^{}_{td} V^*_{tb} \; & =
& \; 0 \; , \nonumber \\
V^{}_{tb} V^*_{ub} + V^{}_{ts} V^*_{us} + V^{}_{td} V^*_{ud} \; & =
& \; 0 \; .
\end{eqnarray}
If $\alpha = 90^\circ$ holds (i.e., both $\triangle^{}_s$ and
$\triangle^{}_c$ are right triangles), then we have ${\rm Re}
(V^{}_{tb} V^{}_{ud} V^*_{td} V^*_{ub}) = 0$ as a straightforward
consequence; namely, the rephasing-invariant quartet $V^{}_{tb}
V^{}_{ud} V^*_{td} V^*_{ub}$ is purely imaginary. Hence $\alpha =
90^\circ$ implies a certain correlation between the parameters of
$V$ in a specific parametrization. Let us illustrate this point by
taking two well-known parametrizations of the CKM matrix $V$.
\begin{itemize}
\item     In the Wolfenstein parametrization of $V$ \cite{W}, we
have
\begin{equation}
{\rm Re}(V^{}_{tb} V^{}_{ud} V^*_{td} V^*_{ub}) \; \approx \;
A^2\lambda^6 \left[ \rho \left( 1 - \rho \right) - \eta^2 \right] \;
.
\end{equation}
So $\alpha = 90^\circ$ coincides with $\eta \approx \sqrt{\rho
\left(1 -\rho\right)}~$ in this parametrization. Taking $\rho
\approx 0.135$ as an example, we obtain $\eta \approx 0.34$. Such
typical values of $\rho$ and $\eta$ are certainly consistent with
current experimental data \cite{PDG08}.

\item     In the standard parametrization of $V$ recommended by
the Particle Data Group \cite{PDG08},
\begin{equation}
{\rm Re}(V^{}_{tb} V^{}_{ud} V^*_{td} V^*_{ub}) \; = \; \cos^2
\theta^{}_{12} \cos^2\theta_{13} \sin\theta^{}_{13}
\cos^2\theta^{}_{23} \left( \tan\theta^{}_{12} \tan\theta^{}_{23}
\cos\delta - \sin\theta^{}_{13} \right) \; .
\end{equation}
Then $\alpha = 90^\circ$ leads to $\cos\delta = \sin\theta^{}_{13}
/\left(\tan\theta^{}_{12} \tan\theta^{}_{23} \right)$. Given
$\theta^{}_{12} \approx 13^\circ$, $\theta^{}_{13} \approx
0.22^\circ$ and $\theta^{}_{23} \approx 2.4^\circ$ for example
\cite{PDG08}, the CP-violating phase turns out to be $\delta
\approx 66^\circ$. This result is also consistent with the
approximate relation $\delta \approx \gamma$ and the present
experimental measurement of $\gamma$ \cite{PDG08}.
\end{itemize}
In both cases, however, we see nothing suggestive behind $\alpha =
90^\circ$.

\vspace{0.4cm}

We proceed to consider a different parametrization of $V$
\cite{FX97}, which is more convenient to explore the underlying
connection between quark masses and flavor mixing angles:
\begin{eqnarray}
V & = & \left( \matrix{ c^{}_{\rm u} & s^{}_{\rm u} & 0 \cr
-s^{}_{\rm u} & c^{}_{\rm u} & 0 \cr 0 & 0 & 1 \cr} \right) \left(
\matrix{e^{-i\phi} & 0 & 0 \cr 0 & c & s \cr 0 & -s& c \cr}
\right) \left( \matrix{ c^{}_{\rm d} & -s^{}_{\rm d} & 0 \cr
s^{}_{\rm d} & c^{}_{\rm d} & 0 \cr 0 & 0 & 1 \cr} \right)
\nonumber \\
& = & \left( \matrix{ s^{}_{\rm u} s^{}_{\rm d} c + c^{}_{\rm u}
c^{}_{\rm d} e^{-i\phi} & s^{}_{\rm u} c^{}_{\rm d} c - c^{}_{\rm
u} s^{}_{\rm d} e^{-i\phi} & s^{}_{\rm u} s \cr c^{}_{\rm u}
s^{}_{\rm d} c - s^{}_{\rm u} c^{}_{\rm d} e^{-i\phi} & c^{}_{\rm
u} c^{}_{\rm d} c + s^{}_{\rm u} s^{}_{\rm d} e^{-i\phi} &
c^{}_{\rm u} s \cr -s^{}_{\rm d} s & -c^{}_{\rm d} s & c \cr}
\right) \; ,
\end{eqnarray}
where $c^{}_{\rm u,d} \equiv \cos\theta^{}_{\rm u,d}$, $s^{}_{\rm
u,d} \equiv \sin\theta^{}_{\rm u,d}$, $c \equiv \cos\theta$ and $s
\equiv \sin\theta$. The merits of this particular parametrization in
understanding quark mass generation and studying heavy flavor
physics are striking \cite{FX97}: (1) it directly follows the chiral
expansion of up- and down-type quark mass matrices, and thus it can
naturally accommodate the observed hierarchy of quark masses; (2)
its three mixing angles are simply but exactly related to the
precision measurements of $B$-meson physics, $\tan\theta^{}_{\rm u}
= |V^{}_{ub}/V^{}_{cb}|$, $\tan\theta^{}_{\rm d} =
|V^{}_{td}/V^{}_{ts}|$ and $\sin\theta = \sqrt{|V^{}_{ub}|^2 +
|V^{}_{cb}|^2}~$; (3) the physical meaning of its mixing angles
$\theta^{}_{\rm u}$ and $\theta^{}_{\rm d}$ can well be interpreted
in a variety of quark mass models (see Ref. \cite{FX00} for a review
with extensive references) with the interesting predictions
$\tan\theta^{}_{\rm u} \approx \sqrt{m^{}_u/m^{}_c}~$ and
$\tan\theta^{}_{\rm d} \approx \sqrt{m^{}_d/m^{}_s}~$; and (4) its
CP-violating phase $\phi$ is closely associated with the light quark
sector, in particular with the mass terms of $u$ and $d$ quarks
\footnote{It is also worth pointing out that this parametrization is
just Euler's three-dimension rotation matrix if the CP-violating
phase $\phi$ is switched off (and a trivial sign rearrangement is
made).}.
Using Eq. (7) to calculate the inner angle $\alpha$ of
$\triangle^{}_s$ and $\triangle^{}_c$, we arrive at
\begin{equation}
\sin \alpha \; = \; \sin\phi \left[ 1 - \left( \tan\theta^{}_{\rm
u} \tan\theta^{}_{\rm d} \cos\theta \cos\phi + \frac{1}{2}
\tan^2\theta_{\rm u} \tan^2\theta_{\rm d} \cos^2\theta + \cdots
\right) \right] \;
\end{equation}
with higher-order terms of $\tan\theta^{}_{\rm u}$ and
$\tan\theta^{}_{\rm d}$ having been omitted. It is clear that
$\alpha \approx \phi$ holds to a good degree of accuracy. Taking
account of $\theta^{}_{\rm u} \approx 5.4^\circ$, $\theta^{}_{\rm
d} \approx 11.5^\circ$ and $\theta \approx 2.4^\circ$ for example
\cite{Gerard}, we obtain either $\alpha \approx 88.9^\circ$ from
$\phi = 90^\circ$ or $\phi \approx 91.1^\circ$ from $\alpha =
90^\circ$. The result $\phi - \alpha \approx 1.1^\circ$ is
interesting in the sense that current experimental data might
imply $\phi = 90^\circ$ at a superhigh energy scale $M_X$ and
$\alpha = 90^\circ$ at the electroweak scale $M^{}_Z$, if
radiative corrections happen to compensate for the tiny
discrepancy between $\alpha (M^{}_Z)$ and $\alpha(M^{}_X)$. We
shall examine whether this point is true or not in the next
section.

\vspace{0.4cm}

Is $\phi$ more fundamental than $\alpha$ in describing the
phenomenon of CP violation in the quark sector? The answer to this
question should be affirmative if the textures of up- and
down-type quark mass matrices ($M^{}_{\rm u}$ and $M^{}_{\rm d}$)
are parallel and originate from the same underlying dynamics
\cite{Xing03}. In this case, $V$ can be decomposed into a product
of two unitary matrices: $V = V^\dagger_{\rm u} V^{}_{\rm d}$,
where $V^{}_{\rm u}$ and $V^{}_{\rm d}$ are responsible
respectively for the diagonalizations of $M^{}_{\rm u}
M^\dagger_{\rm u}$ and $M^{}_{\rm d} M^\dagger_{\rm d}$ (i.e.,
$V^\dagger_{\rm u} M^{}_{\rm u} M^\dagger_{\rm u} V^{}_{\rm u} =
{\rm Diag}\{m^2_{u}, m^2_{c}, m^2_{t}\}$ and $V^\dagger_{\rm d}
M^{}_{\rm d} M^\dagger_{\rm d} V^{}_{\rm d} = {\rm Diag}\{m^2_{d},
m^2_{s}, m^2_{b}\}$) and take the following forms:
\begin{eqnarray}
V^{}_{\rm u} & = & \left( \matrix{e^{-i\phi^{}_x} & 0 & 0 \cr 0 &
c^{}_x & s^{}_x \cr 0 & -s^{}_x & c^{}_x \cr} \right) \left(
\matrix{c^{}_{\rm u} & -s^{}_{\rm u} & 0 \cr s^{}_{\rm u} &
c^{}_{\rm u} & 0 \cr 0 & 0 & 1 \cr} \right) \; , \nonumber \\
V^{}_{\rm d} & = & \left( \matrix{e^{-i\phi^{}_y} & 0 & 0 \cr 0 &
c^{}_y & s^{}_y \cr 0 & -s^{}_y & c^{}_y \cr} \right) \left(
\matrix{c^{}_{\rm d} & -s^{}_{\rm d} & 0 \cr s^{}_{\rm d} &
c^{}_{\rm d} & 0 \cr 0 & 0 & 1 \cr} \right) \; ,
\end{eqnarray}
where $c^{}_{x,y} \equiv \cos\theta^{}_{x,y}$ and $s^{}_{x,y} \equiv
\sin\theta^{}_{x,y}$ are defined. It is obvious that $\theta^{}_y -
\theta^{}_x = \theta$ and $\phi^{}_y - \phi^{}_x = \phi$ hold. Hence
$\phi$ measures the phase difference between up- and down-type quark
mass matrices and is the only source of CP violation in the quark
sector. Let us make a new phenomenological conjecture of the
relationship between $\theta^{}_{x,y}$ (or $\phi^{}_{x,y}$) and
$\theta$ (or $\phi$):
\begin{eqnarray}
\theta^{}_x & = & -Q^{}_{\rm u} \theta \; , ~~~~ \phi^{}_x \;
=\; -Q^{}_{\rm u} \phi \; ; \nonumber \\
\theta^{}_y & = & -Q^{}_{\rm d} \theta \; , ~~~~ \phi^{}_y \; =\;
-Q^{}_{\rm d} \phi \; ,
\end{eqnarray}
where $Q^{}_{\rm u} = +2/3$ and $Q^{}_{\rm d} = -1/3$ are the
electric charges of up- and down-type quarks, respectively. Given
the experimental values of $\theta^{}_{\rm u}$, $\theta^{}_{\rm
d}$, $\theta$ and $\phi$, it is then possible to determine
$V^{}_{\rm u}$ and $V^{}_{\rm d}$ by using Eqs. (9) and (10). The
reconstruction of $M^{}_{\rm u}M^\dagger_{\rm u}$ and $M^{}_{\rm
d}M^\dagger_{\rm d}$ from $V^{}_{\rm u}$ and $V^{}_{\rm d}$ is
straightforward, because the values of six quark masses are all
known \cite{XZZ08}. If both $M^{}_{\rm u}$ and $M^{}_{\rm d}$ are
taken to be Hermitian or symmetric in a particular flavor basis,
then they can directly be reconstructed from quark masses and
flavor mixing parameters.

\section{RGE effects on $\phi$ and $\alpha$}

The one-loop renormalization-group equations (RGEs) of the CKM
matrix elements, together with the RGEs of gauge couplings and the
RGEs of Yukawa couplings of quarks and charged leptons, have
already been calculated by several authors \cite{RGE}. Here we
focus on the RGE running behaviors of $|V^{}_{\alpha i}|^2$ (for
$\alpha =u,c,t$ and $i=d,s,b$) by taking account of $y^2_u \ll
y^2_c \ll y^2_t$ and $y^2_d \ll y^2_s \ll y^2_b$, where
$y^{}_\alpha$ and $y^{}_i$ stand respectively for the eigenvalues
of the Yukawa coupling matrices of up- and down-type quarks. In
this excellent approximation, we simplify the results of Ref.
\cite{RGE} and arrive at
\begin{eqnarray}
& & 16\pi^2 \frac{\rm d}{{\rm d}t} \left[ \matrix{|V^{}_{ud}|^2 &
|V^{}_{us}|^2 & |V^{}_{ub}|^2 \cr |V^{}_{cd}|^2 & |V^{}_{cs}|^2 &
|V^{}_{cb}|^2 \cr |V^{}_{td}|^2 & |V^{}_{ts}|^2 & |V^{}_{tb}|^2
\cr} \right ] \nonumber \\
& = & 2C y^2_b \left[
\matrix{|V^{}_{ud}|^2 |V^{}_{ub}|^2 & |V^{}_{us}|^2 |V^{}_{ub}|^2
& -|V^{}_{ub}|^2 \left( 1 - |V^{}_{ub}|^2 \right) \cr
|V^{}_{td}|^2 |V^{}_{tb}|^2 - |V^{}_{ud}|^2 |V^{}_{ub}|^2 &
|V^{}_{ts}|^2 |V^{}_{tb}|^2 - |V^{}_{us}|^2 |V^{}_{ub}|^2 &
-|V^{}_{cb}|^2 \left( |V^{}_{tb}|^2 - |V^{}_{ub}|^2 \right) \cr
-|V^{}_{td}|^2 |V^{}_{tb}|^2 & -|V^{}_{ts}|^2 |V^{}_{tb}|^2 &
|V^{}_{tb}|^2 \left( 1 -
|V^{}_{tb}|^2 \right) \cr} \right] \nonumber \\
& + & 2C y^2_t \left[ \matrix{|V^{}_{ud}|^2 |V^{}_{td}|^2 &
|V^{}_{ub}|^2 |V^{}_{tb}|^2 - |V^{}_{ud}|^2 |V^{}_{td}|^2 &
-|V^{}_{ub}|^2 |V^{}_{tb}|^2 \cr |V^{}_{cd}|^2 |V^{}_{td}|^2 &
|V^{}_{cb}|^2 |V^{}_{tb}|^2 - |V^{}_{cd}|^2 |V^{}_{td}|^2 &
-|V^{}_{cb}|^2 |V^{}_{tb}|^2 \cr -|V^{}_{td}|^2 \left( 1 -
|V^{}_{td}|^2 \right) & -|V^{}_{ts}|^2 \left( |V^{}_{tb}|^2 -
|V^{}_{td}|^2 \right) & |V^{}_{tb}|^2 \left( 1 - |V^{}_{tb}|^2
\right) \cr} \right] \; ,
\end{eqnarray}
where $t\equiv \ln (\mu /M^{}_Z)$, $C=-1.5$ in the SM and $C=+1$
in the minimal supersymmetric SM (i.e., MSSM). Therefore,
\begin{eqnarray}
&& 16\pi^2 \frac{\rm d}{{\rm d}t} \ln
\frac{|V^{}_{ub}|^2}{|V^{}_{cb}|^2} \; = \; -2C y^2_b \left( 1 -
|V^{}_{tb}|^2 \right) \; , \nonumber \\
&& 16\pi^2 \frac{\rm d}{{\rm d}t} \ln
\frac{|V^{}_{td}|^2}{|V^{}_{ts}|^2} \; = \; -2C y^2_t \left( 1 -
|V^{}_{tb}|^2 \right) \; , \nonumber \\
&& 16\pi^2 \frac{\rm d}{{\rm d}t} \ln \frac{|V^{}_{td}|^2 +
|V^{}_{ts}|^2}{|V^{}_{tb}|^2} \; = \; -2C \left( y^2_b + y^2_t
\right) \; .
\end{eqnarray}
Combining Eqs. (7) and (12), we immediately obtain
\begin{eqnarray}
&& 16\pi^2 \frac{\rm d}{{\rm d}t} \ln\tan\theta^{}_{\rm u} \; =
\; -C y^2_b \sin^2\theta \; , \nonumber \\
&& 16\pi^2 \frac{\rm d}{{\rm d}t} \ln\tan\theta^{}_{\rm d} \; = \;
-C y^2_t \sin^2\theta \; , \nonumber \\
&& 16\pi^2 \frac{\rm d}{{\rm d}t} \ln\tan\theta \; = \; -C \left(
y^2_b + y^2_t \right) \; ;
\end{eqnarray}
or equivalently,
\begin{eqnarray}
&& 16\pi^2 \frac{\rm d \theta^{}_{\rm u}}{{\rm d}t} \; = \;
-\frac{1}{2} C y^2_b \sin 2\theta^{}_{\rm u} \sin^2\theta \; ,
\nonumber \\
&& 16\pi^2 \frac{\rm d \theta^{}_{\rm d}}{{\rm d}t} \; = \;
-\frac{1}{2} C y^2_t \sin 2\theta^{}_{\rm d} \sin^2\theta \; ,
\nonumber \\
&& 16\pi^2 \frac{\rm d \theta}{{\rm d}t} \; = \; -\frac{1}{2} C
\left( y^2_b + y^2_t \right) \sin 2\theta \; .
\end{eqnarray}
Let us stress that the simplicity of RGEs of three quark mixing
angles is naturally expected for our particular parametrization of
$V$, just because its matrix elements involving $t$ and $b$ quarks
are very simple and exactly consistent with the $t$- and
$b$-dominance approximations taken for the RGEs of $|V^{}_{\alpha
i}|^2$ \cite{Xing06}.

\vspace{0.4cm}

We proceed to derive the RGE of the CP-violating phase $\phi$ from
\begin{eqnarray}
16\pi^2 \frac{\rm d}{{\rm d}t} |V^{}_{ud}|^2 & = & 2C
|V^{}_{ud}|^2 \left( y^2_b |V^{}_{ub}|^2 + y^2_t |V^{}_{td}|^2
\right) \nonumber \\
& = & C \sin^2\theta \left( 2\sin^2\theta_{\rm u}
\sin^2\theta^{}_{\rm d} \cos^2\theta + 2\cos^2\theta^{}_{\rm u}
\cos^2\theta^{}_{\rm d} + \sin 2\theta^{}_{\rm u} \sin
2\theta^{}_{\rm d} \cos\theta \cos\phi \right) \nonumber
\\
& & \times \left( y^2_b \sin^2\theta^{}_{\rm u} + y^2_t
\sin^2\theta^{}_{\rm d} \right) \; .
\end{eqnarray}
Note that the derivative of $|V^{}_{ud}|^2$ can be given in terms
of the derivatives of $\theta^{}_{\rm u}$, $\theta^{}_{\rm d}$,
$\theta$ and $\phi$ as follows:
\begin{eqnarray}
\frac{\rm d}{{\rm d}t} |V^{}_{ud}|^2 & = & \left[ \sin 2
\theta^{}_{\rm u} \left( \sin^2\theta^{}_{\rm d} \cos^2\theta -
\cos^2\theta^{}_{\rm d} \right) + \cos 2\theta^{}_{\rm u} \sin
2\theta^{}_{\rm d} \cos\theta \cos\phi
\right] \frac{\rm d \theta^{}_{\rm u}}{{\rm d} t} \nonumber \\
& + & \left[ \sin 2 \theta^{}_{\rm d} \left( \sin^2\theta^{}_{\rm
u} \cos^2\theta - \cos^2\theta^{}_{\rm u} \right) + \sin
2\theta^{}_{\rm u} \cos 2\theta^{}_{\rm d} \cos\theta \cos\phi
\right] \frac{\rm d \theta^{}_{\rm d}}{{\rm d} t} \nonumber \\
& - & \left[ \sin^2 \theta^{}_{\rm u} \sin^2\theta^{}_{\rm d} \sin
2\theta + \frac{1}{2} \sin 2\theta^{}_{\rm u} \sin 2\theta^{}_{\rm
d} \sin\theta \cos\phi \right] \frac{\rm d \theta}{{\rm d} t}
\nonumber \\
& - & \left[ \frac{1}{2} \sin 2\theta^{}_{\rm u} \sin
2\theta^{}_{\rm d} \cos\theta \sin\phi \right] \frac{\rm d
\phi}{{\rm d} t} \; .
\end{eqnarray}
Substituting Eqs. (14) and (15) into the right- and left-hand
sides of Eq. (16), respectively, we simply arrive at
\begin{equation}
16\pi^2 \frac{\rm d \phi}{{\rm d} t} \; = \; 0 \; .
\end{equation}
This result implies that the CP-violating phase $\phi$ is stable
against radiative corrections at the one-loop level and in the
approximation of quark mass hierarchies (i.e., $y^2_u \ll y^2_c
\ll y^2_t$ and $y^2_d \ll y^2_s \ll y^2_b$). With the help of Eqs.
(14) and (17), a straightforward calculation of the derivative of
$\alpha$ given in Eq. (8) leads to
\begin{equation}
16\pi^2 \frac{\rm d \alpha}{{\rm d} t} \; = \; 0 \; .
\end{equation}
Hence the RGE running effect of $\alpha$ is also negligibly small,
implying that the low-energy result $\phi - \alpha \approx
1.1^\circ$ essentially keeps unchanged even if $\mu \gg M^{}_Z$
holds. In other words, it is impossible to get $\alpha = 90^\circ$
at $M^{}_Z$ from $\phi = 90^\circ$ at $M^{}_X$ through the
one-loop RGE evolution.

\vspace{0.4cm}

Such a conclusion remains valid at the two-loop level. By using
the two-loop RGEs of the CKM matrix elements \cite{RGE2}, we have
carried out a numerical analysis of the running behaviors of
$\phi$ and $\alpha$ from $M^{}_X$ to $M^{}_Z$ (or vice versa) in
both the SM and the MSSM
\footnote{H. Zhang and S. Zhou did this numerical exercise for me.
Their RGE program has also been used to evaluate the running
masses of quarks and leptons at different energy scales
\cite{XZZ08}.}.
Here are our main observations: (1) the RGE running effect of
$\phi$ or $\alpha$ is too small (less than $0.1^\circ$ from
$M^{}_Z \sim 10^2 ~{\rm GeV}$ to $M^{}_X \sim 10^{16} ~ {\rm
GeV}$) in the SM or in the MSSM with $\tan\beta
> 1.5$; and (2) it cannot compensate for the small phase difference
$\phi - \alpha \approx 1.1^\circ$ no matter how we adjust the
energy scale (and the value of $\tan\beta$ in the MSSM case).

\section{Quark-lepton complementarity}

Compared with the parametrization of the CKM matrix $V$ given in Eq.
(7), a similar parametrization of the MNS matrix $U$ is also
convenient for the description of lepton flavor mixing and CP
violation
\footnote{This parametrization may naturally arise from the parallel
(and probably hierarchical) textures of charged-lepton and neutrino
mass matrices. It is phenomenologically possible to obtain
$\theta^{}_l \approx \arctan \left(\sqrt{m^{}_e/m^{}_\mu} ~\right)
\approx 4^\circ$ together with a suggestive relationship
$\theta^{}_\nu \approx \arctan \left(\sqrt{m^{}_1/m^{}_2}~\right)$
\cite{FX06}, where $m^{}_1$ and $m^{}_2$ are the neutrino masses
corresponding to $\nu^{}_e$ and $\nu^{}_\mu$ flavors. Furthermore,
$U$ can be decomposed into $U = U^\dagger_l U^{}_\nu P$ in a way
similar to Eqs. (9) and (10) with $Q^{}_l = -1$ and $Q^{}_\nu =0$.}:
\begin{eqnarray}
U & = & \left ( \matrix{ c^{}_l & s^{}_l   & 0 \cr -s^{}_l    &
c^{}_l   & 0 \cr 0   & 0 & 1 \cr } \right )  \left ( \matrix{
e^{-i\varphi}  & 0 & 0 \cr 0   & c & s \cr 0   & -s    & c \cr }
\right )  \left ( \matrix{ c^{}_{\nu} & -s^{}_{\nu}  & 0 \cr
s^{}_{\nu} & c^{}_{\nu}   & 0 \cr 0   & 0 & 1 \cr } \right ) P
\nonumber \\
& = & \left ( \matrix{ s^{}_l s^{}_{\nu} c + c^{}_l c^{}_{\nu}
e^{-i\varphi} & s^{}_l c^{}_{\nu} c - c^{}_l s^{}_{\nu}
e^{-i\varphi} & s^{}_l s \cr c^{}_l s^{}_{\nu} c - s^{}_l
c^{}_{\nu} e^{-i\varphi} & c^{}_l c^{}_{\nu} c + s^{}_l s^{}_{\nu}
e^{-i\varphi} & c^{}_l s \cr - s^{}_{\nu} s   & - c^{}_{\nu} s   &
c \cr } \right ) P \; ,
\end{eqnarray}
where $c^{}_{l,\nu} \equiv \cos\vartheta^{}_{l,\nu}$,
$s^{~}_{l,\nu} \equiv \sin\vartheta_{l,\nu}$, $c \equiv
\cos\vartheta$ and $s \equiv \sin\vartheta$; and $P$ is a diagonal
phase matrix containing two nontrivial CP-violating phases when
three neutrinos are Majorana particles. Although the form of $U$
in Eq. (19) is apparently different from that of the standard
parametrization of $U$ \cite{PDG08}, their corresponding flavor
mixing angles ($\vartheta^{}_l, \vartheta^{}_\nu, \vartheta$) and
($\vartheta^{}_{12}, \vartheta^{}_{13}, \vartheta^{}_{23}$) have
quite similar meanings in interpreting the experimental data on
solar and atmospheric neutrino oscillations. In the limit
$\vartheta^{}_l = \vartheta^{}_{13} = 0$, one can easily arrive at
$\vartheta^{}_\nu = \vartheta^{}_{12}$ and $\vartheta =
\vartheta^{}_{23}$. Note that the tri-bimaximal neutrino mixing
pattern \cite{TB}, which is well consistent with a global fit of
current neutrino oscillation data \cite{FIT}, {\it does} coincide
with this interesting limit (i.e., $\vartheta^{}_l =
\vartheta^{}_{13} = 0^\circ$, $\vartheta^{}_\nu =
\vartheta^{}_{12} = \arctan(1/\sqrt{2}) \approx 35.3^\circ$ and
$\vartheta = \vartheta^{}_{23} =45^\circ$). Therefore, three
mixing angles of $U$ can simply be related to those of solar,
atmospheric and reactor neutrino oscillations in the leading-order
approximation \cite{Xing06}; i.e., $\vartheta^{}_{\rm sol} \approx
\vartheta^{}_\nu$, $\vartheta^{}_{\rm atm} \approx \vartheta$ and
$\vartheta^{}_{\rm rea} \approx \vartheta^{}_l \sin\vartheta$ as a
natural consequence of very small $\vartheta^{}_l$.

\vspace{0.4cm}

The above comparison between our parametrization and the standard
one indicates that both of them might be suitable for describing
the approximate quark-lepton complementarity (QLC) relations
\cite{Raidal}. The latter means the following empirical
observations in the standard parametrizations of the CKM and MNS
matrices:
\begin{equation}
\theta^{}_{12} + \vartheta^{}_{12} \; \approx \; 45^\circ \; ,
~~~~ \theta^{}_{23} + \vartheta^{}_{23} \; \approx \; 45^\circ \;
,
\end{equation}
where $\theta^{}_{ij}$ and $\vartheta^{}_{ij}$ (for $1 \leq i < j
\leq 3$) represent quark and lepton mixing angles, respectively.
Eq. (20) is actually consistent with the present experimental data
within $1\sigma$ error bars \cite{PDG08}. Turning to our
parametrizations of the CKM and MNS matrices in Eqs. (7) and (19),
we find that similar QLC relations can approximately hold within
$1\sigma$ error bars:
\begin{equation}
\theta^{}_{\rm d} + \vartheta^{}_\nu \; \approx \; 45^\circ \; ,
~~~~~~~~~~ \theta + \vartheta \; \approx \; 45^\circ \; .
\end{equation}
This result seems to be somewhat contrary to the expectation that
the QLC relations are convention-dependent and may only hold in a
single parametrization for $V$ and $U$ \cite{J2}. We believe that
the {\it exact} QLC relations can only be realized (or assumed) in
a unique parametrization for $V$ and $U$, but the approximate ones
are possible to show up in different parametrizations. The reason
for the latter point is quite simple: the smallest elements of the
CKM and MNS matrices are both at their up-right corner (i.e.,
$|V^{}_{ub}| = \sin\theta^{}_{13} = \sin\theta^{}_{\rm u}
\sin\theta = \cdots$ and $|V^{}_{e3}| = \sin\vartheta^{}_{13} =
\sin\vartheta^{}_l \sin\vartheta = \cdots$), and thus the flavor
mixing between the first and second families is approximately
decoupled from that between the second and third families. In
other words, it is the smallness of $\theta^{}_{13}$ (or
$\theta^{}_{\rm u}$) and $\vartheta^{}_{13}$ (or $\vartheta^{}_l$)
that assures the approximate QLC relations in Eqs. (20) and (21)
to hold simultaneously.

\vspace{0.4cm}

Note again that the approximate QLC relations, similar to $\alpha
\approx 90^\circ$, are extracted from current experimental data at
low energies. One may wonder whether such empirical relations are
stable against radiative corrections, or whether they can be exact
at a specific energy scale far above $M^{}_Z$. Because quark and
lepton flavor mixing angles obey different RGEs in their evolution
from $M^{}_Z$ to $M^{}_X$ (or vice versa) \cite{RGE3}, we should
have ${\rm d}\theta^{}_{12}/{\rm d}t + {\rm
d}\vartheta^{}_{12}/{\rm d}t \neq 0$ and ${\rm
d}\theta^{}_{23}/{\rm d}t + {\rm d}\vartheta^{}_{23}/{\rm d}t \neq
0$ in general \cite{Xing05,Schmidt}. This observation is also true
for our parametrizations of the CKM and MNS matrices (see Ref.
\cite{Xing06} for the explicit RGEs of $\vartheta^{}_l$,
$\vartheta^{}_\nu$, $\vartheta$ and $\varphi$), no matter whether
neutrinos are Dirac particles or Majorana particles.

\vspace{0.4cm}

Finally, let us conjecture that $\varphi = 90^\circ$ holds in the
lepton sector. This possibility can actually be realized in some
specific neutrino mass models (e.g., $\varphi = 90^\circ$ was first
obtained in the so-called ``democratic" neutrino mixing scenario
\cite{FX96}). While $\phi$ is rather stable against quantum
corrections from one energy scale to another, as already shown in
Eq. (18), $\varphi$ is in general sensitive to the RGE effects
\cite{Xing06}. Does $\varphi = 90^\circ$ imply that a pair of the
leptonic unitarity triangles are right or almost right? The answer
to this question depends on the value of $\vartheta^{}_l$ (or
equivalently $\vartheta^{}_{13}$ in the standard parametrization of
$U$), which has not been fixed by current neutrino oscillation
experiments. For illustration, we consider the leptonic unitarity
triangle $\triangle^{}_1$ defined by the orthogonality relation
$V^{}_{e2} V^*_{e3} + V^{}_{\mu 2} V^*_{\mu 3} + V^{}_{\tau 2}
V^*_{\tau 3} = 0$ in the complex plane \cite{FX00}. Denoting the
inner angle $\alpha^{}_l \equiv \arg[-(V^{}_{\mu 2} V^*_{\mu
3})/(V^{}_{e2} V^*_{e3})]$ and taking the maximal CP-violating phase
$\varphi = 90^\circ$, we find
\begin{equation}
\sin\alpha^{}_l \; = \; 1 - \frac{1}{2} c^2_l s^2_l c^{-2}_\nu
s^{-2}_\nu c^{-2} \left(s^2_\nu - c^2_\nu c^2\right)^2 + \cdots \; ,
\end{equation}
where higher-order terms of $s^{}_l$ have been omitted. Then
$\alpha^{}_l \approx 89.5^\circ$ can be obtained from Eq. (22)
with the typical inputs $\vartheta^{}_l \approx 5^\circ$,
$\vartheta^{}_\nu \approx 34^\circ$ and $\vartheta \approx
45^\circ$. It is easy to see that the value of $\alpha^{}_l$
approaches $\varphi = 90^\circ$ when $\vartheta^{}_l$ approaches
zero, but in the limit of $\vartheta^{}_l = 0^\circ$ there will be
no CP violation (i.e., $\varphi$ becomes trivial and can be
rotated away from $U$ by rephasing the electron field) and all the
leptonic unitarity triangles of $U$ must collapse into lines. This
example illustrates that $\varphi = 90^\circ$ implies the
existence of two nearly right unitarity triangles
($\triangle^{}_1$ and its counterpart $\triangle^{}_\tau$ defined
by the orthogonality relation $V^{}_{e1} V^*_{\mu 1} + V^{}_{e2}
V^*_{\mu 2} + V^{}_{e3} V^*_{\mu 3} =0$) in the lepton sector,
similar to the case in the quark sector.

\section{Summary and concluding remarks}

In view of the experimental indication that two unitarity triangles
of the CKM matrix $V$ are almost the right triangles with $\alpha
\approx 90^\circ$, we have explored its possible implications on the
phenomenology of quark flavor mixing and quark-lepton
complementarity. Taking account of a very suggestive parametrization
of $V$, we have shown that its CP-violating phase $\phi$ is nearly
equal to $\alpha$ (i.e., $\phi - \alpha \approx 1.1^\circ$). Both
$\phi$ and $\alpha$ are stable against the renormalizaton-group
evolution from the electroweak scale $M^{}_Z$ to a superhigh energy
scale $M^{}_X$ or vice versa, and thus it is impossible to obtain
$\alpha = 90^\circ$ at $M^{}_Z$ from $\phi = 90^\circ$ at $M^{}_X$.
We have conjectured that there might also exist a maximal
CP-violating phase $\varphi \approx 90^\circ$ in our parametrization
of the MNS matrix $U$. The approximate quark-lepton complementarity
relations, which hold in the standard parametrizations of $V$ and
$U$ (i.e., $\theta^{}_{12} + \vartheta^{}_{12} \approx 45^\circ$ and
$\theta^{}_{23} + \vartheta^{}_{23} \approx 45^\circ$), can also
hold in our particular parametrizations of $V$ and $U$ (i.e.,
$\theta^{}_{\rm d} + \vartheta^{}_\nu \approx 45^\circ$ and $\theta
+ \vartheta \approx 45^\circ$). We have pointed out that the reason
for this interesting coincidence simply comes from the smallness of
$|V^{}_{ub}|$ and $|V^{}_{e3}|$.

\vspace{0.4cm}

At this point, it is worthwhile to remark that the
phenomenological ansatz proposed in Eq. (10) can be elaborated on
so as to obtain an explicit texture of quark mass matrices. A
similar ansatz can be made for the lepton sector by adopting the
parametrization of $U$ advocated in Eq. (19) and decomposing it
into $U = U^\dagger_l U^{}_\nu P$ in a way exactly analogous to
Eqs. (9) and (10) with $Q^{}_l = -1$ and $Q^{}_\nu = 0$. For
simplicity, here we only illustrate how to reconstruct the
Hermitian quark mass matrices $M^{}_{\rm u}$ and $M^{}_{\rm d}$ by
using Eqs. (9) and (10). After taking account of the smallness of
three mixing angles and the hierarchy of six quark masses, we
approximately arrive at
\begin{eqnarray}
M^{}_{\rm u} & \approx & \left( \matrix{\lambda^{}_u +
\lambda^{}_c \theta^2_{\rm u} & -\lambda^{}_c \theta^{}_{\rm u}
e^{+i 60^\circ} & \displaystyle -\frac{2}{3} \lambda^{}_c
\theta^{}_{\rm u} \theta e^{+i 60^\circ} \cr\cr -\lambda^{}_c
\theta^{}_{\rm u} e^{-i 60^\circ} & \displaystyle \lambda^{}_c +
\frac{4}{9} \lambda^{}_t \theta^2 & \displaystyle -\frac{2}{3}
\lambda^{}_t \theta \cr\cr \displaystyle -\frac{2}{3} \lambda^{}_c
\theta^{}_{\rm u} \theta e^{-i 60^\circ} & \displaystyle
-\frac{2}{3} \lambda^{}_t \theta &
\lambda^{}_t \cr} \right) \; , \nonumber \\
M^{}_{\rm d} & \approx & \left( \matrix{\lambda^{}_d +
\lambda^{}_s \theta^2_{\rm d} & -\lambda^{}_s \theta^{}_{\rm d}
e^{-i 30^\circ} & \displaystyle +\frac{1}{3} \lambda^{}_s
\theta^{}_{\rm d} \theta e^{-i 30^\circ} \cr\cr -\lambda^{}_s
\theta^{}_{\rm d} e^{+i 30^\circ} & \displaystyle \lambda^{}_s +
\frac{1}{9} \lambda^{}_b \theta^2 & \displaystyle +\frac{1}{3}
\lambda^{}_b \theta \cr\cr \displaystyle +\frac{1}{3} \lambda^{}_s
\theta^{}_{\rm d} \theta e^{+i 30^\circ} & \displaystyle
+\frac{1}{3} \lambda^{}_b \theta & \lambda^{}_b \cr} \right) \; ,
\end{eqnarray}
where $|\lambda^{}_q| = m^{}_q$ (for $q=u,c,t$ and $d,s,b$),
$\theta^{}_{\rm u} \approx 9.4 \times 10^{-2}$, $\theta^{}_{\rm d}
\approx 2.0 \times 10^{-1}$ and $\theta \approx 4.2 \times
10^{-2}$. Such a parallel texture of up- and down-type quark mass
matrices is certainly suggestive and may serve as a
phenomenological starting point of model building. For instance,
setting $(M^{}_{\rm u})^{}_{11} = (M^{}_{\rm d})^{}_{11} = 0$
leads to two interesting relations $\theta^{}_{\rm u} \approx
\sqrt{m^{}_u/m^{}_c}~$ and $\theta^{}_{\rm d} \approx
\sqrt{m^{}_d/m^{}_s}~$.

\vspace{0.4cm}

Although different parametrizations of the CKM matrix $V$ are
mathematically equivalent, one of them might be able to make the
underlying physics of quark flavor mixing more transparent and to
establish simpler connections between the observable quantities
and the model parameters. We find that our parametrization of $V$
in Eq. (7) {\it does} satisfy the above criterion. We expect that
the similar parametrization of the MNS matrix $U$ in Eq. (19) is
also useful in describing lepton flavor mixing. Needless to say,
much more experimental, phenomenological and theoretical attempts
are desirable in order to solve three fundamental flavor puzzles
in particle physics --- the generation of fermion masses, the
dynamics of flavor mixing and the origin of CP violation.

\vspace{0.6cm}

The author would like to thank H. Zhang and S. Zhou for their
helps in running the two-loop RGE program. This work was supported
in part by the National Natural Science Foundation of China under
grant No. 10425522 and No. 10875131, and in part by the Ministry
of Science and Technology of China under grant No. 2009CB825207.

\begin{figure*}[t]
\centering
\vspace{2cm}
\includegraphics[width=110mm]{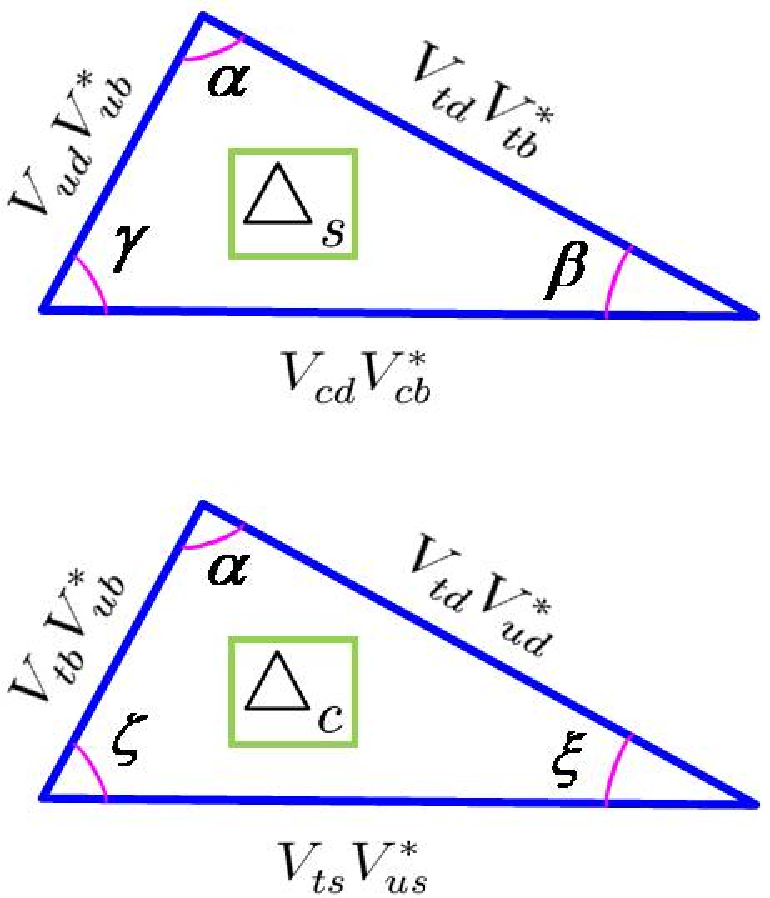}
\vspace{0.7cm} \caption{The CKM unitarity triangles $\triangle^{}_s$
and $\triangle^{}_c$ defined in the complex plane.}
\end{figure*}

\end{document}